\documentclass[preprint,pra,showpacs,showkeys]{revtex4}
\usepackage{graphicx,amsmath,amsfonts,amssymb}
\usepackage{times}

\begin{document}

\title{Entanglement and quantum phase transition in  alternating XY spin
chain with next-nearest neighbour interactions \footnote{Supported
by the Key Higher Education Programme of Hubei Province under Grant
No Z20052201, the Natural Science Foundation of Hubei Province,
China under Grant No 2006ABA055, and the Postgraduate Programme of
Hubei Normal University under Grant No 2007D20.}}

\author{C. J. Shan\footnote{ E-mail: scj1122@163.com}}
\author{W. W. Cheng}
\author{T. K. Liu\footnote{Corresponding author. E-mail:
tkliuhs@163.com}}
\author{Y. X. Huang}
\author{H. Li}
\affiliation{College of Physics and Electronic Science, Hubei Normal
University, Huangshi 435002, China}
\date{\today}

\begin{abstract}

By using the method of density-matrix renormalization-group to solve
the different spin-spin correlation functions, the
nearest-neighbouring entanglement(NNE) and next-nearest-neighbouring
entanglement(NNNE) of one-dimensional alternating Heisenberg XY spin
chain is investigated in the presence of alternating nearest
neighbour interactions of exchange couplings, external magnetic
fields and next-nearest neighbouring interactions. For dimerized
ferromagnetic spin chain, NNNE appears only above the critical
dimerized interaction, meanwhile, the dimerized interaction effects
quantum phase transition point and improves NNNE to a large value.
We also study the effect of ferromagnetic or antiferromagnetic
next-nearest neighboring (NNN) interactions on the dynamics of NNE
and NNNE. The ferromagnetic NNN interaction increases and shrinks
NNE  below and above critical frustrated interaction respectively,
while the antiferromagnetic NNN interaction always decreases NNE.
The antiferromagnetic NNN interaction results to a larger value of
NNNE in comparison to the case when the NNN interaction is
ferromagnetic.
\end{abstract}

\pacs{03.65.Ud, 03.67.Mn, 75.10.Pq}

\keywords{the entanglement; alternating XY spin chain; next-nearest
neighbour interactions}

\maketitle

\section{Introduction}
\indent  Entanglement is one of the most profound features of
quantum mechanics and has been considered as an important resource
in quantum information processing (QIP) including
teleportation,$^{[1]}$ cryptography and secure direct
communication,$^{[2,3]}$ and quantum communication and
computation.$^{[4]}$  In order to realize quantum information
process, great effort has been devoted to the generation of
entanglement in linear optics,$^{[5]}$ cavity QED $^{[6-8]}$ and ion
trap$^{[9,10]}$ schemes. In recent years much attention has been
focused on the spin system with Heisenberg exchange interaction,
which is a typical quantum system, and the entanglement properties
have been extensively investigated in spin systems, such as Ising
model$^{[11]}$ and isotropic and anisotropic Heisenberg
models.$^{[12-15]}$ However; as far as we know, most discussions
mentioned above merely focused on the models with nearest neighbour
interaction of exchange couplings, while next-nearest neighbour
interaction has not been taken into account. Dimerized systems and
frustrated systems with next-nearest-neighbour (NNN)
interactions$^{[16-18]}$ play an important role in condensed matter
theory, and entanglement in dimerized and frustrated systems has
been considered.$^{[19-23]}$ Sun et al$^{[19]}$ investigate
entanglement properties in dimerized and frustrated spin-one models
by applying the concept of negativity. In Ref.[20], entanglement is
studied in an open alternating chain of nuclear spins $s = 1/2$ with
spin-spin couplings in an external magnetic field under the
thermodynamic equilibrium conditions. Gu et al$^{[21]}$ have
investigated entanglement in frustrated spin-half Heisenberg chains.
Chen Yan et al$^{[22]}$ analyzed sublattice entanglement and quantum
phase transitions in antiferromagnetic spin chains, Chen Shu et
al$^{[23]}$ study the fidelity and quantum phase transition for the
Heisenberg chain with next-nearest-neighbouring interaction. In
fact, some one-dimensional and two-dimensional compounds
($CuGeO_{3}, NaV_{2}O_{5}$) have manifested such interactions.
Therefore, it is worthwhile to include next-nearest neighbouring
interaction in the
studies of spin chain entanglement.\\
\indent The 1D Heisenberg model is a simple but realistic and
extensively studied in solid state system. Recently, Osterloh et al
$^{[24]}$ examined the entanglement between two spins of position i
and j in the spin chains for the pure case, Huang et al,$^{[25]}$
Osenda etal$^{[26]}$ and we$^{[27]}$ have demonstrated that
entanglement can be controlled  by introducing impurities into the
systems. Meanwhile, the entanglement shares many features with
quantum phase transition (QPT) for a many-body system.$^{[28,29]}$
QPT, which occurs at absolute zero temperature and is purely driven
by quantum fluctuation, is the structural change in the properties
of the ground state.  The associated level crossings lead to the
presence of non-analyticities in the energy spectrum. Therefore, the
knowledge about the ground-state entanglement, the nonlocal
correlation in quantum systems, is considered as a key to further
understand QPT. In this paper, we study the pairwise ground state
entanglement between the nearest-neighbor sites and that of the
next-nearest neighbouring sites in one-dimensional $s=\frac{1}{2}$
Heisenberg XY spin chain with dimerised exchange couplings and
next-nearest neighbour coupling, to our knowledge, which has not
been reported before. The present study will help us to further
understand the behavior of the entanglement and QPT in
one-dimensional alternating Heisenberg XY model with next-nearest
neighbour interactions, and this model can display a variety of
interesting physical phenomena.  More interestingly, we can control
or manipulate the NNE and NNNE in QPT point with the help of
dimerised exchange couplings, dimerised external magnetic fields and
frustrated exchange interaction.
\section{Solution of the XY model and spin-spin correlation functions}
\indent We consider one-dimensional alternating Heisenberg XY model
of N spin-$\frac{1}{2}$ particles with nearest-neighbour
interactions and next-nearest neighbour interactions.
One-dimensional Hamiltonian can be written as
\begin{eqnarray}
H=&-&\frac{1+\gamma}{2}\sum_{i=1}^{N}J_{i,i+1}\sigma _{i}^{x}\sigma
_{i+1}^{x}-\frac{1-\gamma }{2}\sum_{i=1}^{N}J_{i,i+1}\sigma
_{i}^{y}\sigma _{i+1}^{y}-\sum_{i=1}^{N}h_{i}\sigma
_{i}^{z}\nonumber\\&-&\frac{1+\gamma
}{2}\sum_{i=1}^{N}J_{i,i+2}\sigma _{i}^{x}\sigma
_{i+2}^{x}-\frac{1-\gamma }{2}\sum_{i=1}^{N}J_{i,i+2}\sigma
_{i}^{y}\sigma _{i+2}^{y}
\end{eqnarray}
where $J_{i,i+1}$ and $J_{i,i+2}$ are alternating exchange
interaction and nearest-neighbour interactions respectively, $h_{i}$
is the strength of external magnetic field on site i,
$\sigma^{x,y,z}$ are the Pauli matrices, $\gamma$ is a dimensionless
parameter characterizing the anisotropy of the model and N is the
total number of sites. Furthermore, the periodic boundary conditions
satisfy $\sigma _{N+1}^{x}=\sigma _{1}^{x}, \sigma _{N+1}^{y}=\sigma
_{1}^{y}, \sigma _{N+1}^{z}=\sigma
_{1}^{z}$.\\
\indent The spectrum of this Hamiltonian can be determined exactly
by a straightforward application of the standard methods. The first
step in the procedure is to perform a Jordan-Wigner
transformation$^{[30]}$ by introducing Fermi
 operators $c_{j}^{+}$, $c_{j}$, as a result, the Hamiltonian (1) is mapped in the free fermion Hamiltonian
\begin{eqnarray}
H=&-&\sum_{i=1}^{N}J_{i,i+1}[(c_{i}^{+}c_{i+1}+\gamma
c_{i}^{+}c_{i+1}^{+})+h.c]-2\sum_{i=1}^{N}h_{i}(c_{i}^{+}c_{i}-\frac{
1}{2})\nonumber\\&-&\sum_{i=1}^{N}J_{i,i+2}[(c_{i}^{+}c_{i+2}+\gamma
c_{i}^{+}c_{i+2}^{+})+h.c]
\end{eqnarray}
In this paper,  the exchange interaction and external magnetic field
have the form $J_{2i-1,2i}=J_{1}=J$, $J_{2i,2i+1}=J_{2}=\alpha J$,
$h_{2i,2i+1}=h_{2i-1,2i}(h)=\beta h$, $i=1,2,\cdots, \frac{N}{2})$
respectively, where $\alpha=J_{2}/ J_{1}$ introduces the dimerised
parameter, the ratio of next-nearest ($J_{i,i+2}=J_{3}$) to
nearest-neighbor interaction coefficients $\kappa=J_{3}/ J_{1}$ is
called the frustration parameter.  $J<0$ and $J>0$ correspond to the
antiferromagnetic and the ferromagnetic cases, respectively.  For
$J_{3}=0$, Eq. (2) reduces to one-dimensional alternating(dimerised)
Heisenberg model, for $J_{1}=J_{2}$, it reduces to one-dimensional
frustrated Heisenberg model.  \\
\indent The density-matrix renormalization-group
(DMRG)method$^{[31,32]}$is applied to obtain the spin correlation
function. The two-point reduced density matrix obtained by tracing
the full density matrix of the system over all sites except the pair
$\{i, j\}$ has the form
\begin{eqnarray}
\rho_{i,j}=\frac{1}{4}(I_{i,j}+\langle\sigma^{z}_{i}\rangle\sigma^{z}_{i}+\langle\sigma^{z}_{j}\rangle\sigma^{z}_{j}+
\sum_{k=x,y,z}\langle\sigma^{k}_{i}\sigma^{k}_{j}\rangle\sigma^{k}_{i}\sigma^{k}_{j})
\end{eqnarray}
In this part, we choose the concurrence defined by Wootters
$^{[33]}$ as a measurement of the pairwise entanglement.
 For a pure or mixed state
of two qubits described by the density matrix $\rho$, the
concurrence C may be calculated explicitly as
\begin{eqnarray}
C(\rho)=\max (0,\lambda _{1}-\lambda _{2}-\lambda _{3}-\lambda _{4})
\end{eqnarray}
where $\lambda _{1}$, $\lambda _{2}$, $\lambda _{3}$, $\lambda _{4}$
are the eigenvalues in descending order of the spin-flipped density
operator R, which is defined by
 $R=\sqrt{\sqrt{\rho }\tilde{\rho} \sqrt{\rho }}$
, where $\tilde{\rho} =(\sigma _{y}\otimes \sigma _{y})\rho ^{\ast
}(\sigma _{y}\otimes \sigma _{y})$, $\tilde{\rho}$ denotes the
complex conjugate of $\rho$, $\sigma _{y}$ is  the usual Pauli
matrix. All the matrix elements in the density matrix can be
calculated from the different spin-spin correlation functions.

\section{Results and discussions}
In this paper,  we focus our discussion on the transverse Ising
model with $\gamma=1$ and introduce the dimensionless parameter
$\lambda=J/2h$. The goal of our present study is to find dynamic
characteristics of the entanglement in varying the dimerised
parameter of exchange couplings, external magnetic fields and NNN
interaction. Figure 1 displays the diagram of the
$J_{1}-J_{2}-J_{3}$ model which can be considered either as a linear
chain(a) or a zigzag chain(b). The alternating nearest-neighbouring
interaction is $J_{1}$ and $J_{2}$ respectively, while the
next-nearest-neighbouring interaction is parameterized by $J_{3}$.
We consider three different physical models, i.e, dimerized
ferromagnetic spin chain, frustrated
ferromagnetic-ferromagnetic(F-F) spin chain, frustrated
ferromagnetic-antiferromagnetic(F-AF) spin chain.  First, we examine
the dimerized Heisenberg chain, this model is characterized by an
alternation of strong and weak bonds between two
nearest-neighbouring spins. Figure 2 shows the nearest neighbouring
concurrence C(1,2) and C(2,3), which correspond to two
nearest-neighbouring spins coupled by bonds $J_{1}$ and $J_{2}$
respectively, as a function of the reduced coupling constant
$\lambda$ at different values of the dimerised parameter $\alpha$ of
exchange couplings, parameter $\beta$ of external magnetic fields
with the system size N =59. Numerical results in Figs.2(a) and 2(c)
show that the concurrence C(1,2) decreases when the dimerised
parameter $\alpha$ of exchange couplings increases, while for the
concurrence C(2,3) the situation is opposite, and the pair of qubits
(1,2) and (2,3) have the same pairwise entanglement for the
homogeneous chain $\alpha=1$. In fact, in the case of  weak
alternation of the nearest-neighbor exchange coupling, we have here
the dimerised spin chain which can be considered qualitatively as a
set of non-interacting spin pairs. In the limit of strong
dimerization, the concurrence reaches a maximum value. Moreover,
spin 2 can be entangled both with spin 1 and with spin 3. Since spin
2 is strongly entangled with spin 1, the entanglement of spins 2 and
3 is weaker. Obviously, the increase of exchange interaction $J_{2}$
suppresses the entanglement of spins coupled by $J_{1}$. The
increase of $J_{2}$ can enhance the entanglement between spins
coupled by $J_{2}$. This explains the behavior of the concurrence
displayed in Fig.2. The effect of the alternation of external
magnetic field $\beta$ is also shown in Figs.2(b) and 2(d). However,
different from the effect of the exchange couplings, the concurrence
(1,2) and (2,3) have similar behavior and increase firstly and then
decrease with increasing the value of the parameter $\beta$,
suggesting a critical alternation of external magnetic field
$\beta_{c}$, where the maximum concurrence occurs, must be exist.\\
\indent From Fig.2, we can see that the alternate interaction plays
an important role in enhancing the nearest neighbouring concurrence.
The maximum value of the concurrence between neighboring sites does
not occur at the critical point. The reasons are based on the
properties of shared entanglement to expect that this maximum should
occur away from the critical point. Entanglement sharing is relevant
to the quantum phase transition in the transverse Ising model as it
provides a fundamental bound on the amount of entanglement that may
be distributed among the other sites, which means that as the
overall entanglement in the lattice is increased, some sites become
disentangled. However, as we all know, the maximum value of the
concurrence between the next-nearest-neighbouring sites does occur
at the critical point $\lambda=1$, so it is necessary to discuss in
this subsection the effect on the next-nearest-neighbouring
entanglement(NNNE).  Figure 3(a) and 3(b) show the change of
concurrence C(1,3) as a function of $\lambda$ for different values
of alternate interactions $\alpha$ and $\beta$ with $\kappa=0$, i.e.
in the absence of NNN interaction. We can see that there is no
entanglement at  $\alpha=0.5$ and $\beta=0.5$. The weak alternate
interactions $\alpha$ and $\beta$ suppress and finally completely
destroy the NNN concurrence. The dashed line in Fig.3(a) depicts at
the critical point, $\lambda=1$, of the transverse Ising
model($\alpha=1$), which maximum next-nearest-neighbouring
entanglement occurs, there is a fundamental transition in the
structure of the ground state. This is consistent with the former
result. As $\alpha$ increases the NNN concurrence tends to increase
faster, the peak value induced by the NN interaction increases and
the $\lambda_{m}$, where concurrence approaches a maximum, shifts to
left very rapidly. The dimerized interaction does assist the
entanglement of formation because of the inside alternation.
Therefore, we can further understand the relation between the
entanglement and quantum transition. In Figs.3(c) and 3(d), we
plotted our numerical results for the threshold NN alternate
interactions $\alpha$ and $\beta$ for the case of
next-nearest-neighboring concurrence. Only above the critical  $\alpha$ and $\beta$, NNN  concurrence appears.\\
\indent For the frustrated model, there is a competition between the
NN and NNN interactions. So the model displays some special features
of entanglement. In Fig.4, we give the results of the NN and NNN
concurrence as a function of the parameter $\lambda$ for different
frustrated interaction $\kappa$. The NNN interaction has a different
frustration effect on C(1,2). For frustrated
ferromagnetic-ferromagnetic(F-F) spin chain $\kappa>0$, by the
comparison among the different curves in Fig.4(a), it is interesting
to find that the entanglement peak between the nearest neighbours in
the dotted line  increase to a value larger than those in the solid
line. With the increasing of $\kappa$, in the dash-dotted line, the
concurrence decreases. We can imagine there must be a critical
frustration strength ($\kappa_{c}$), below $\kappa_{c}$, NNN
interaction enhances entanglement, while above $\kappa_{c}$, NNN
interaction shrinks entanglement. This behaviour is due to the
energy level cross at the point $\kappa = 0.5$, as seen clearly from
Fig.4(a). Before and after the critical point, the entanglements
display distinct behaviours. However, for frustrated
ferromagnetic-antiferromagnetic(F-AF) spin chain $\kappa<0$, the
situation becomes different due to the frustration, with the
increase in absolute value of $\kappa$ and weak parameter $\lambda$,
antiferromagnetic NNN interaction always shrinks entanglement, which
is different from the results obtained from Fig.4(a). Another
important character revealed by Fig.4(b) is that, as parameter
$\lambda$ increases, the concurrence increases and tends to move to
infinity($>0$) by increasing the value of the parameter $\kappa$,
the strong $\kappa$ is helpful to keeping the better entanglement.
It is the frustrated interaction that leads to considerable
different evolutions of the entanglement, hence the entanglement is
rather sensitive to any change with the NNN interaction. From
Figs.4(c) and 4(d), we can see that the height of the peak increases
with the increase of antiferromagnetic $\kappa$, which is a
characteristic feature of one dimensional Heisenberg
antiferromagnetic chain. This suggests that the antiferromagnetic
component becomes significant for larger $\kappa$. Meanwhile, the
antiferromagnetic NNN interaction results to a larger value of NNNE
in comparison to the case when the NNN interaction is ferromagnetic.
In order to explain these results it is necessary to take into
account the influence of the competing roles played by NN and NNN
interaction on the entangled states. Thus, by adjusting NNN
interaction constant, one can control or manipulate the NN and NNN
entanglement.\\
\indent In summary, from the above analysis, it is clearly noted
that dimerised exchange couplings, dimerised external magnetic
fields and frustrated exchange interaction, which play the competing
roles in enhancing quantum entanglement, have a notable influence on
NN and NNN concurrence in one-dimensional $s=\frac{1}{2}$  XY spin
system. The NN and NNN concurrence exhibits some interesting
phenomena. For dimerized ferromagnetic spin chain, NNNE appears only
above the critical dimerized interaction, which effects quantum
phase transition point and increases NNNE. For a
ferromagnetic-antiferromagnetic frustrated spin chain, the NNN
interaction is predominant in the competing role and can enhance the
entanglement to a steady value. The ferromagnetic NNN interaction
increases and shrinks NNE below and above critical frustrated
interaction respectively, while the antiferromagnetic NNN
interaction always decreases NNE. Meanwhile, the antiferromagnetic
NNN interaction can generate a larger value of NNNE in comparison to
the case when the NNN interaction is ferromagnetic. So we can employ
NNN interaction strength as well as dimerized interaction to realize
quantum entanglement control. As for the case of $\gamma\neq1$(XY
model), we will present further reports in the future.
\\

\newpage

\newpage
\begin{figure}
\begin{center}
\includegraphics[width=1.0\textwidth]{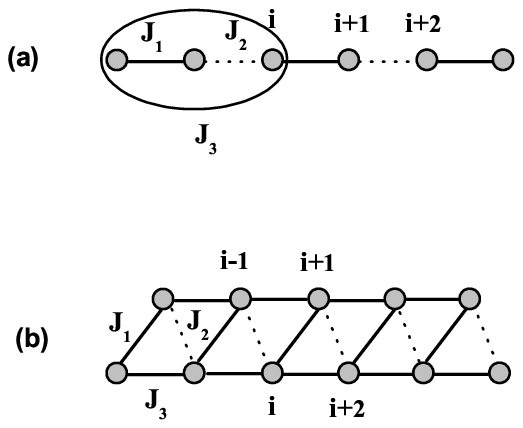}\\
\caption{Diagram of the $J_{1}-J_{2}-J_{3}$ model which can be
considered either as a linear chain(a) or a zigzag chain(b). The
alternating nearest-neighbouring interaction is $J_{1}$ and $J_{2}$
respectively, while the next-nearest-neighbouring interaction is
parameterized by $J_{3}$}
\end{center}
\end{figure}
\newpage
\begin{figure}
\begin{center}
\includegraphics[width=1.0\textwidth]{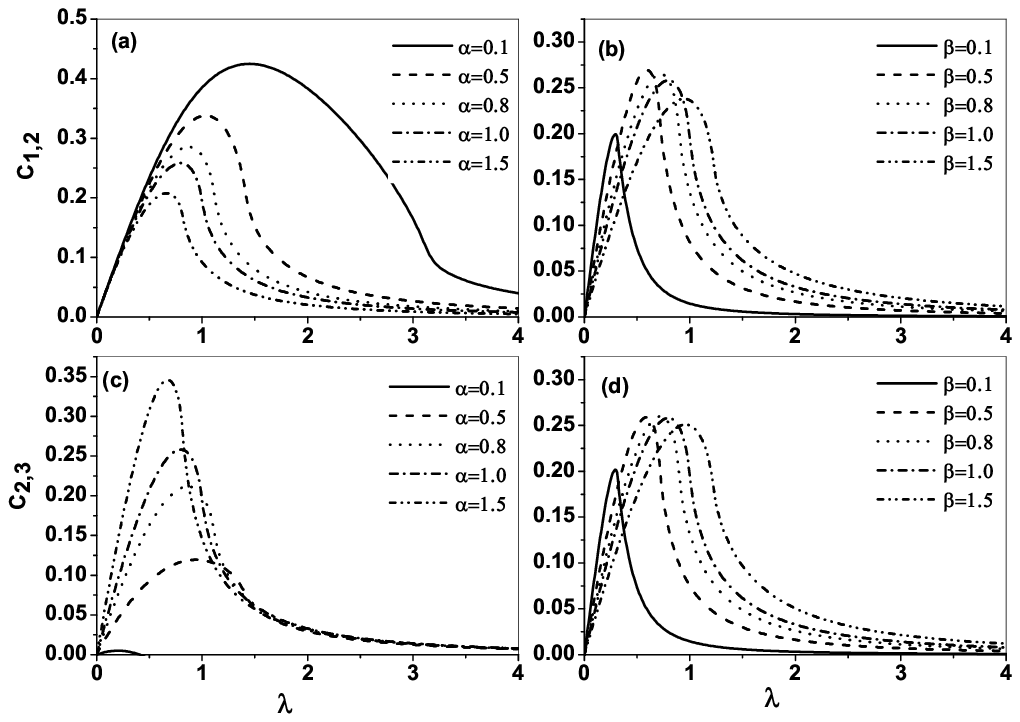}\\
\caption{ The nearest neighbouring concurrence $C_{1,2}$ and
$C_{2,3}$ as a function of the reduced coupling constant $\lambda$
for various alternating values of the nearest-neighbour interaction
and external magnetic field, NNN interaction $\kappa=0$.}
\end{center}
\end{figure}
\newpage
\begin{figure}
\begin{center}
\includegraphics[width=1.0\textwidth]{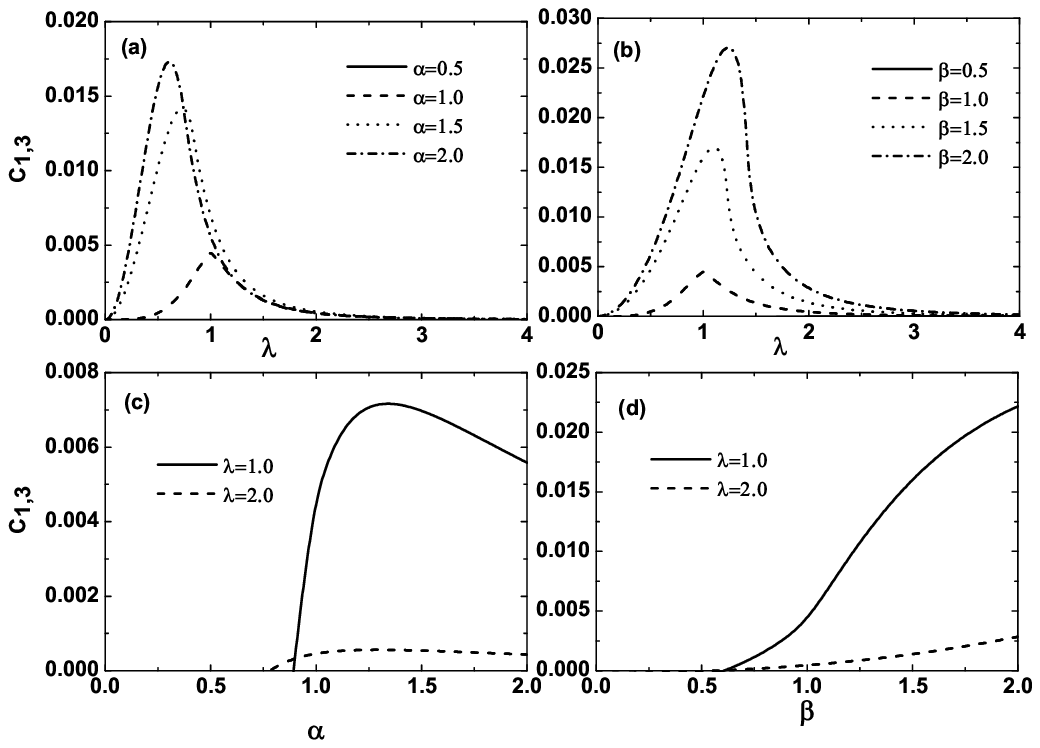}\\
\caption{The next-nearest-neighbouring concurrence $C_{1,3}$ as a
function of the reduced coupling constant $\lambda$ for various
alternating values of the nearest-neighbour interaction and external
magnetic field, NNN interaction $\kappa=0$. }
\end{center}
\end{figure}
\newpage
\begin{figure}
\begin{center}
\includegraphics[width=1.0\textwidth]{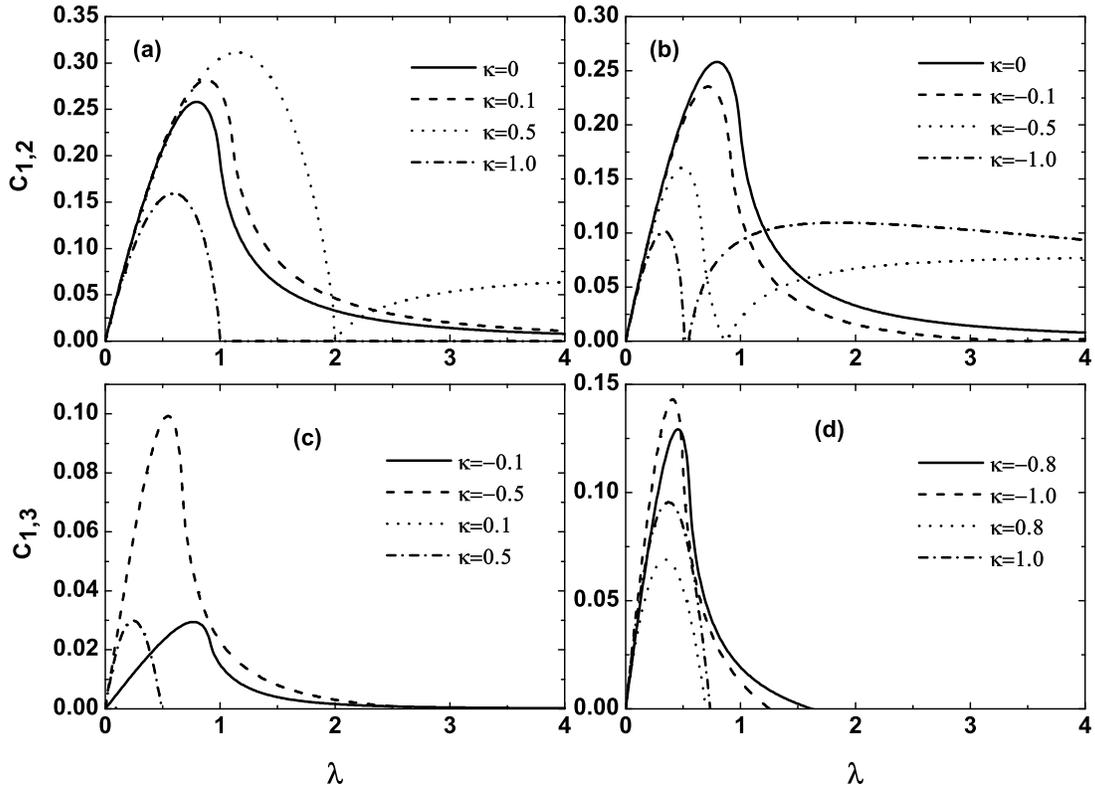}\\
\caption{The concurrence $C_{1,2}$ and $C_{1,3}$ as a function of
the reduced coupling constant $\lambda$ for various frustrated
values of next-nearest-neighbouring interaction, NN interaction
$\alpha=\beta=1.0$.}
\end{center}
\end{figure}
\end{document}